\documentstyle[preprint,eqsecnum,aps,epsfig]{revtex}
\tightenlines
\newcommand{\cs}{\'{c}}

\newcommand{\beq}{\begin{equation}}
\newcommand{\eeq}{\end{equation}}
\newcommand{\bdm}{\begin{displaymath}}
\newcommand{\edm}{\end{displaymath}}
\newcommand{\beqa}{\begin{eqnarray}}
\newcommand{\eeqa}{\end{eqnarray}}
\newcommand{\beqab}{\begin{eqnarray*}}
\newcommand{\eeqab}{\end{eqnarray*}}

\def\nn{\nonumber}

 \def\@makefnmark{\hbox to 0pt{$^{\@thefnmark}$\hss}}  %ORIGINAL

\newcounter{saveeqn}%

\begin{document}

%%%%%%%%
\draft
%%%%%%%

%\begin{flushright}
%NYU-TH-02-03-11 \\

%\end{flushright}

%%%%%%%%%%%%%%%%%%%%%%%%%
\preprint{NYU-TH-03-01-01}
%%%%%%%%%%%%%%%%%%%%%%%%%

\title{Gravity induced over a smooth soliton}

\author{ 
Marko Kolanovi\cs \footnote{e-mail: mk679@nyu.edu} }
%%%%%%%%%%%%%%%%%%%%%%%%%%%%%%%%%%%%%%%%%%%%%%%%%%%%%%%%%%
\address{Center for Cosmology and Particle Physics\\ Department of Physics, New York University,
New York, NY 10003}
%%%%%%%%%%%%%%%%%%%%%%%%%%%%%%%%%%%%%%%%%%%%%%%%%%%%%%%%%

\date{\today}
\maketitle
\begin{abstract}

I consider gravity induced 
over a smooth (finite thickness) soliton. Graviton kinetic term is coupled to
bulk scalar that develops solitonic vacuum expectation value. 
Couplings of Kaluza-Klein modes to soliton-localized matter are suppressed, giving rise to 
crossover distance $r_c=M_{P}^2/M_{*}^3$ between 4D and
5D behavior. This system can be viewed as a finite  thickness brane 
regularization  of the model of Dvali, Gabadadze and Porrati.

\end{abstract}

%%%%%%%%%%%%%
%\narrowtext
%%%%%%%%%%%%
\newpage

\section{Introduction}
\setcounter{equation}{0}

Brane worlds theories with large or infinite extra dimensions recently
provided insight to a number of problems of high energy physics and 
possible relations between them. Those models can emerge as a part of
the fundamental higher dimensional theory described in terms of strings.
Even if extra dimensions are not realized in nature, they are valuable 
testing ground for new ideas in particle physics and gravity.

For some time, it was believed that the extra dimensions can exist only if
they are compact or have a finite volume. The problem with the infinite volume extra
dimensions was how to obtain a four dimensional force of gravity between objects
located at the brane. In work \cite{dgp}, it was shown that the four dimensional
Newton's force arises on the brane due to an induced kinetic term. 
The model of Dvali, Gabadadze and Porrati \cite{dgp} is described by the action 

\beq\label{DGP}
S=M_{*}^3\int d^{4}xdy\sqrt{|G|}{\cal R}_{5}+M_{P}^2\int d^{4}x dy\delta(y)\sqrt{|g|}R_{4}.
\eeq
Here $y$ is the coordinate of an extra dimension, ${\cal R}_{5}$ is 5D curvature, $G_{MN}$ 5D metric,
$R_4$ is 4D curvature constructed from metric induced on the brane $g_{\mu\nu}\equiv
G_{MN}(y=0)\delta^{M}_{\mu}\delta^{N}_{\nu}$, $M_*$ is the fundamental scale of gravity
and $M_P$ scale that characterizes brane rigidity. In order to obtain the correct
value for Newton's constant on the brane, $M_P$ has to be taken $\sim 10^{19}$GeV.
The brane is represented with the delta function i.e. thickness of the brane is taken to be zero. 
A related model with one infinite extra dimensions was presented and studied in \cite{rele}.
Phenomenology of the action (\ref{DGP}) was studied in \cite{dgkn}. There it was shown that the
scale of the bulk gravity can be as low as $M_{*}\sim 10^{-3}$eV. Further properties of the model, 
cosmological evolution, new manifestations at astronomical
scales and string theory realizations were examined in \cite{furt,dis,str}. 

In these notes, I study a similar system of induced (enhanced) gravity over a smooth soliton
\beq\label{SYS}
S=M_{*}^{3}\int d^{4}xdy\sqrt{|G|}{\cal R}_{5}(1+\frac{M_{P}^3}{M_{*}^3} \chi^{2}(M_{P}y)).
\eeq
Here $\chi(M_{P}y)$ is a localized function representing a smooth tension-less soliton. I 
find that at the linearized level, features of (\ref{DGP}) and (\ref{SYS}) are the
same. In short, four dimensional gravity at distance $r<r_c\equiv M_{P}^2/M_{*}^{3}$ is mediated by
a metastable massless resonance. Width of the resonance is $\sim r_{c}^{-1}$. At distances $r>r_c$,
gravity becomes five dimensional. Therefore I conclude that (\ref{SYS}) is a finite thickness
brane regularization of (\ref{DGP}).

Through the paper I will work with the simpler 'scalar gravity' model that captures
main features of gravity.
A Variation of the strength of the kinetic term can be a consequence of dilaton-like coupling to 
bulk scalara ($\chi$) that exhibits solitonic vev. I neglect the gravitational effect of the soliton,
which is equivalent to fine-tunning its tension to zero.
In the next section I study properties of the spectrum of Lagrangian in which the
strength of kinetic term is a function of the extra coordinate. Results are applicable for
scalar and gauge bulk field as well as for linearized gravity. In section 3, I investigate the 
spectrum of Kaluza Klein modes in system (\ref{SYS}) and their coupling to soliton-localized matter.
I find the crossover distance $r_c=M_{P}^2/M_{*}^{3}$, both from 
the suppression of couplings of KK modes 
to matter localized on soliton and from an equivalent problem of tunneling through 
the potential barrier. I find no new scale  $\sim\sqrt{r_c/a}$ (where
$1/a$ is the brane thickness), contrary to results of  
\cite{kiri}, where the different regularization scheme was considered. 
I conclude that the system (\ref{SYS}) can be viewed as a finite thickness brane regularization of
action (\ref{DGP}). In the appendix I outline the procedure for finding the spectrum
of fermionic modes when the kinetic term of a bulk fermion field is varying through
the extra dimension.

%%%%%%%%%%%%%%%%%%%%%%%%%%%%%%%%%%%%%%%%%%%%%%%%%%%%%%%%%%%%%%%%%%%%%%%%%%%%%%%%%%%%%%%%%

\section{Localization in model with non-homogeneous kinetic term} 
\setcounter{equation}{0}

In this section I investigate general properties of a model in which 
the strength of a kinetic term for a bulk field is not homogeneous through
the infinite extra dimension. 
Let us take that the bulk field $\Phi(x_{\mu},y)$ 
is coupled through the kinetic term to a field $\chi(x_{\mu},y)$

\beq\label{solmod}
{\cal L}=f(\chi)\partial_{A}\Phi\partial^{A}\Phi+\partial_{A}\chi\partial^{A}\chi
-V(\chi).
\eeq
Function $f(y)$ characterizes the coupling of two fields.
Field $\chi(x_{\mu},y)$ has potential $V(\chi)$ and can obtain an expectation
value in the form of a soliton $\chi\longrightarrow \chi_{Cl}(y)$. Then, the Lagrangian
for the field $\Phi$ has the following form

\beq\label{firstl}
{\cal L}=f(y)\partial_{A}\Phi\partial^{A}\Phi.
\eeq
Field $\Phi$ can represent  bulk scalar, gauge field or graviton in linearized approximation. 

To find the four dimensional spectrum of (\ref{firstl}) we decompose field $\Phi$
into Kaluza Klein modes $\Phi(x_{\mu},y)=\sum_{m}\Phi_{m}(y)\sigma_{m}(x_{\mu})$, where
$\Box\sigma_{m}=-m^2\sigma_{m}$. The differential equation for wave functions $\Phi_{m}(y)$ is

\beq\label{ekv}
f(y)\partial_{y}^2\Phi_{m}+\partial_{y}f(y)\partial_{y}\Phi_{m}+f(y)m^2\Phi_{m}=0.
\eeq
Taking equations (\ref{ekv}) for two different eigenvalues $m,n$, cross-multiplying them with
eigenfunctions, subtracting and integrating over the length of extra dimension 
we get the orthogonality relation

\beq\label{orth}
\int f(y)\Phi_{n}\Phi_{m}dy=\alpha_{m}\delta_{m,n},
\eeq
where $\alpha_{m}$ is the normalization constant. Combining (\ref{ekv}) and (\ref{orth}) we get 

\beq\label{nogo}
\frac{\int f(y)\partial_{y}\Phi_{n}\partial_{y}\Phi_{m}dy}{\int f(y)\Phi_{n}\Phi_{m}dy}=m^2.
\eeq
For non-negative $f(y)$, both the denominator and numerator of this expression are positive, which
means that there are no ghosts, tachions or ghost tachions in effective four dimensional
action obtained from (\ref{firstl}).  We can introduce additional coupling of field $\Phi$ to a 
'charge' density  $\Psi(y)\rho(x)$ localized at the soliton (centered at $y=0$). 
In 4D effective theory the coefficient of a coupling is determined by the convolution
of the matter profile $\Psi(y)$ and the wave function $\Phi_{m}$
(an average of wave function over the soliton-localized matter) 

\beq\label{convo}
\tilde{\Phi}_{m}(0)\equiv
\int \Psi(y)\Phi_{m}(y)dy.
\eeq
Effective action for  modes $\sigma_{m}$ then reads

\beq\label{efac}
{\cal L}=\frac{1}{2}\sum_{m}\left( \partial_{\mu}\sigma_{m}\partial^{\mu}\sigma_{m}-m^2\sigma_{m}^{2}
\right)+\sum_{m}\frac{\tilde{\Phi}_{m}(0)}{\sqrt{M_{*}^{3}\alpha_{m}}}\sigma_{m}\rho(x),
\eeq
where $M_{*}$ is the fundamental scale. Zero mode $\sigma_{0}$
will be normalizable for choices of $f(y)$ whose integral is finite, while it will be a single
mode in continuum for functions $f(y)$ that are nonzero at $y\longrightarrow \pm\infty$.
This feature distinguishes between the cases of localization and quasi-localization.
In order to find the spectrum and determine localization properties of KK modes, one has to
specify the function $f(y)$. To better understand the connection between the shape of $f(y)$ and 
localization properties, it is useful to transform (\ref{ekv}) to related Schroedinger problem

\beq\label{schro}
\Phi_{m}\equiv\frac{\phi_m}{\sqrt{f}},\quad -\partial_{y}^2\phi_{m}+V(y)\phi_{m}=m^2\phi_{m},
\eeq
where the localizing potential $V$ has the form

\beq\label{locpo}
V(y)=\left( \frac{1}{2}\frac{f''(y)}{f(y)}-\frac{1}{4}\left(\frac{f'(y)}{f(y)}\right)^2\right).
\eeq
Whether this potential will behave as an attractive well or a repulsive barrier depends on the
balance between the magnitude of the second and first derivative of the profile $f(y)$.

\subsection{$f(y)=e^{-y^2a^2}$}

Now we consider two choices of function $f(y)$ that lead to the localization
of part or all of the spectrum of KK excitations. 
First example is a Gaussian profile $f(y)=e^{-y^2a^2}$.
Here, $a$ is the scale that determines the brane width.
This profile will give rise to a localized tower of KK modes with masses
$m\sim\sqrt{n}$, where $n$ is the integer. This 
was used as a model of localization of gauge fields in \cite{dv}. Here we repeat it because
we will use it as a starting point for the investigation of soliton induced 
gravity in the next section.

Potential (\ref{locpo}) that results from this profile is the potential of a simple
harmonic oscillator shifted by a constant so that the zero point energy vanishes

\beq\label{sho}
V(y)=a^2(a^2y^2-1).
\eeq
Normalized wave functions (\ref{schro}) are 

\beq\label{wfho}
\phi_{n}(y)=\frac{1}{\sqrt{2^{n}n!\pi^{1/2}}}e^{-(ya)^2/2}H_{n}(ya),
\eeq
where the spectrum of masses goes like

\beq\label{spho}
m_{n}=a\sqrt{2n},\quad n=0,1,2,...
\eeq
The interaction of those modes with some charge located at the brane is determined
by the value of the wave function at the position of the charge. For example, coupling of a charge
at $y=0$  is

\beq\label{coupho}
{\cal L}_{I}=\sum_{n}\frac{(-)^{n/2}\sqrt{an!}}{2^{n/2}\sqrt{M_{*}^3}(n/2)!}\sigma_{n}\rho,
\eeq
for $n$ even, while it vanishes for odd $n$. The vanishing of coupling for odd values
of $n$ is also true for a smooth brane that localize charge $\rho$. The reason is
that wave functions of modes with odd $n$ are odd, while the matter localization profile 
$\Psi(y)$ is even and coupling (\ref{convo}) vanishes.

\subsection{$f(y)=e^{-|ya|}$}

Other example is the exponentially decaying profile $f(y)=e^{-|ya|}$. This is presented
as an example of a  localized profile $f(y)$ with a finite integral, that
doesn't localize all of the KK modes (as one could naively expect).
The Schroedinger potential for this profile is

\beq\label{pot2}
V(y)=\frac{1}{4}a^2-a\delta(y).
\eeq
Spectrum of modes consist of a single localized zero mode

\beq\label{lzm}
\phi_{0}=\sqrt{a/2}e^{-a|y|/2},
\eeq
and a continuum of scattering states starting at  $m>a/2$. The potential between two charges
$q_1$ and $q_2$ located at $y=0$ due to exchange of field $\Phi$ would change from a four dimensional
law to a five dimensional law at distance $r\sim 1/a$.

\section{Smooth brane regularization of DGP model}
\setcounter{equation}{0}

In this section I establish the relation between the model of Dvali, Gabadadze and Porrati and
action (\ref{SYS}) at linearized level.
Let us consider the five dimensional model with two scales. Fundamental scale $M_*$ determines
the action of the bulk filed $\Phi$. The potential for the other bulk field $\chi$
is governed by the much higher brane scale $a$. Therefore, scale $a$ 
determines the width of the brane, which is soliton of field $\chi$. Field 
$\chi$ is coupled quadratically to a kinetic term of $\Phi$, and the coupling is
suppressed in powers of fundamental scale $M_*$

\beq\label{pc}
{\cal L}=\left(1+\frac{\chi^2}{M_{*}^3}\right)\partial_{A}\Phi\partial^{A}\Phi+
\partial_{A}\chi\partial^{A}\chi-V(\chi).
\eeq
Let us take that the field $\chi$ gets expectation 
value $\chi_{Cl}=a^{3/2}\exp(-(ya)^2/2)$ so that the profile function (\ref{firstl}) is

\beq\label{dgpr}
f(y)=1+\left(\frac{a}{M_{*}}\right)^3e^{-(ya)^2}\equiv 1+Ae^{-(ya)^2}.
\eeq
Since we take that the soliton scale $a$ is much bigger than the bulk scale $M_*$,
coefficient $A$ is a huge number. The Schroedinger potential (\ref{locpo}) for this 
profile is 

\beq\label{potq}
V(y)=-Aa^2\frac{A(1-(ya)^2)+e^{(ya)^2}(1-2(ya)^2)}{(A+e^{(ya)^2})^2}.
\eeq
Potential (\ref{potq}) is shown in Fig. 1 together with the potential of harmonic oscillator
(\ref{sho}).

\vspace{5mm}
\centerline{\epsfig{file=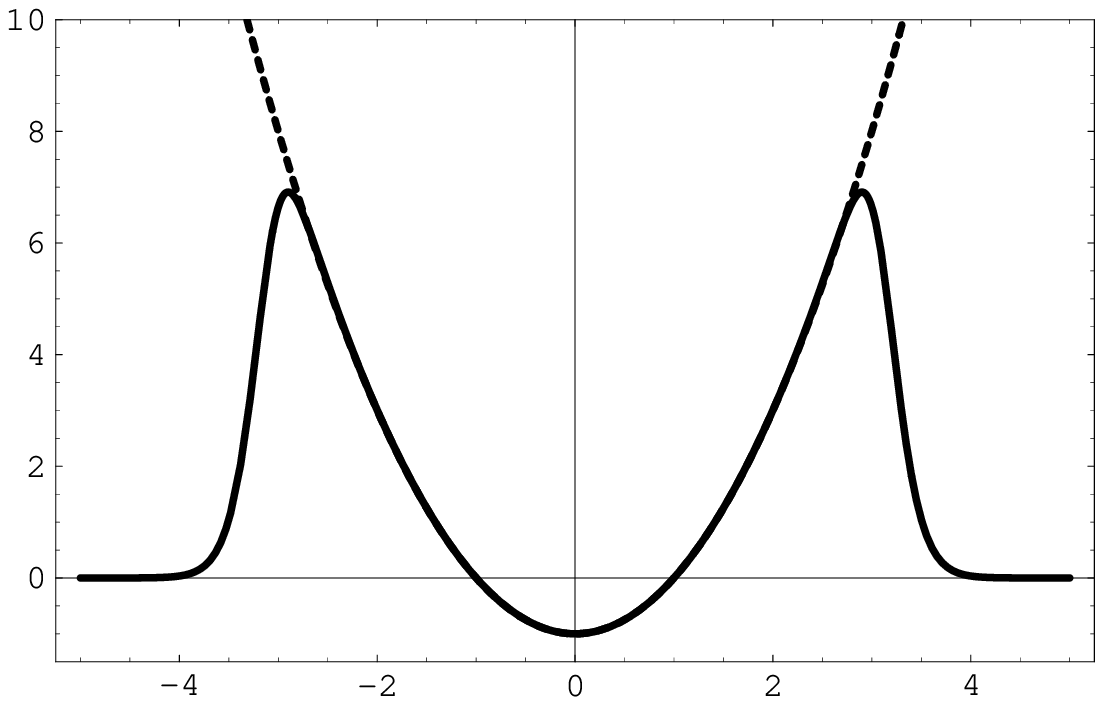,width=10cm}}
{\footnotesize\textbf{Figure 1:} Potential (\ref{potq}) (solid line)
together with the harmonic oscillator potential
(\ref{sho}) (dashed line). $y$ is in units of inverse $a$, the potential is in units of $a^2$,
while $A=10000$.}
\vspace{5mm}

For small values of $y$, the potential coincides with the potential of a harmonic oscillator. After
certain a distance $\sim 1/a$, the square well is 'cut' and the 
potential rapidly drops to zero. From this
behavior we can qualitatively predict the spectrum of states. Lowest bound states of the harmonic
oscillator will obtain width and will tunnel through the barrier. Since we are interested
in the case of gravity we want to find the decay width of a zero mode that will mediate
the four dimensional Newton's force on the brane. 
Higher modes have mass $\sim a$ and are irrelevant for low
energy physics. Decay width in the WKB approximation for the zero mode is

\beq\label{decw}
\Gamma_{0}\sim a\exp\left( -2\int_{y_1}^{y_2}\sqrt{V}dy\right),
\eeq
where $y_1,y_2$ are classical turning points and $1/a$ is the width of the well. 
For large values of $A$ condition for WKB validity $|V'/(2V)^{3/2}|\ll 1$ is well satisfied
everywhere (except at classical turning points). The integral in
(\ref{decw}) can be evaluated numerically for the 
large range of values $A$. The integral is independent
of $a$ and behaves as $\sim A^{0.488}$ (power has negligible numerical error). Therefore,
in the WKB approximation zero mode has a decay width

\beq\label{zmdw}
\Gamma_{0}\sim\frac{a}{A^{0.976}}.
\eeq
Wave functions of the continuous Kaluza-Klein spectrum can be found by solving the
Schroedinger equation. We find that the couplings of modes (\ref{convo}) to matter
localized on the soliton are suppressed. 

Here, we can explore the question of the universality of the coupling of filed $\Phi$ (scalar gravity)
to brane matter.
If different types of matter localized on the soliton have different profiles $\Psi(y)$, coupling
(\ref{convo}) of gravity would be non-universal. This is sometimes argued to be a problem
\footnote{non-universality of coupling can in fact be a desirable property of models with infinite
extra dimensions. It opens the possibility for solving the cosmological constant problem without
spoiling general covariance of the resulting four dimensional theory. This was pointed out to me
by Gia Dvali.} for
theories with gravity induced over a smooth brane \cite{rub}. We adopt the following framework
in which the violation of universality is suppressed by soliton thickness scale $a$.
The potential for a scalar field that forms solitonic brane
is governed by high mass scale $a$. Here we take that $a\sim M_P$. Mass of Standard Model 
fields is much lower than the cutoff $a$. We take that SM particles come from the 
zero mode of another bulk field, while the mass splitting is due to some higher order effect. 
Zero modes have profiles roughly of a soliton itself $\Psi_{0}\sim a\exp(-(ya)^2)$. Potential well
that localizes zero mode can have bound states whose masses would be $\sim a$. Indeed 
wave functions of those states $\Psi_{B.S.}$ are different and lead to the different 
coupling (\ref{convo}). 
However, because of their large mass they are decoupled from low energy physics. 
Now let us return to the spectrum of potential (\ref{potq}).
The Schroedinger equation for (\ref{potq}) can be solved numerically.
For simplicity, we take that the matter profile is $\Psi_{0}\sim a\exp(-(ya)^2)$ and evaluate coupling
(\ref{convo}) for the continuum of KK modes. The qualitative behavior of wave functions is
as follows. Outside the well, wave functions are plane waves of frequency $m$. Inside the
well, wave functions are roughly Gaussians with mass-dependent height and constant width
($\sim 1/a$). Approximately, one can take for the coupling 
$\tilde{\Psi}_{m}(0)\approx \sqrt{\pi}\Psi_{m}(0)$. 
Actual coupling was evaluated for each mode numerically.
Dependence of coupling (\ref{convo}) on the mass of KK mode is shown in Fig.2

\vspace{5mm}
\centerline{\epsfig{file=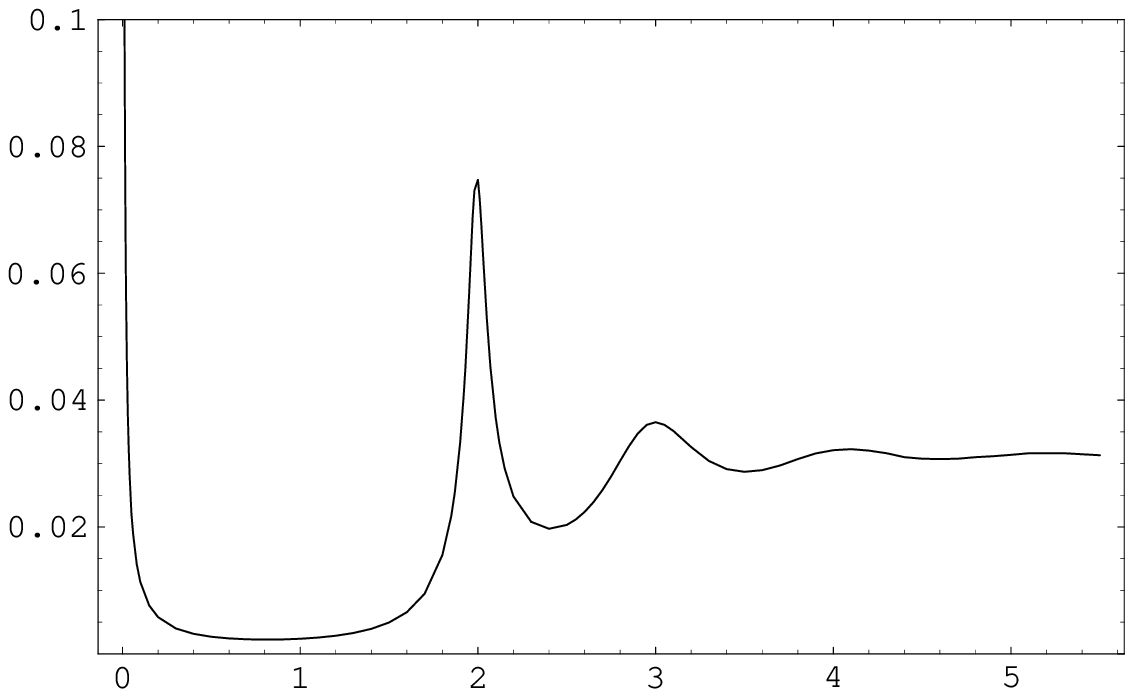,width=12cm}}
{\footnotesize\textbf{Figure 2:} Suppression of KK coupling for $A=10000$. 
For $m=0$ coupling has value 1 (outside the graph range). Peaks are at positions of resonant 
states of harmonic oscillator $m=0,2,\sqrt{8}...$~. Log-log plot
of the same graph (multiplied with $\sqrt{A}$) is shown in the inset. 
The solid line has coefficient -1 to show
the suppression of coupling  $\sim 1/m$.}
\vspace{5mm}

There is a peak located at the mass zero. This is the zero mass graviton resonance that is responsible
for the 4D force at short distances. This resonance starts at coupling 1 and decays
as $\sim 1/m$ (see the log-log plot).
Dependence of the half-width of the peak at $m=0$ on the magnitude of $A$ can be evaluated by 
numerical analysis. By sampling data for the large range of values $A$
($10^2<A<10^5$) we find

\beq\label{wi}
\Delta m \sim \frac {a}{A^{0.997\pm 0.005}}.
\eeq
Besides the metastable 'massless' graviton there is no other
resonances until we get to higher states of the harmonic oscillator that have mass $\sim a$.
In particular, nothing unusual
is happening at scale  $\sim a/\sqrt{A}$ contrary to the results of \cite{kiri} 
where a different regularization was used (there advocated mass scale
$a/\sqrt{A}$ would be on position -4.605 on the log-log graph in Fig.2).

Newtonian potential due to an exchange of field $\Phi$ between two masses
at short distances is given by

\beq\label{newt}
V(r)\approx \int_{0}^{\infty}\frac{|\tilde{\Phi}_{m}(0)|^{2}}
{M_{*}^3}\frac{e^{-mr}}{r}dm\approx\frac{a}{M_{*}^{3}A}
\frac{1}{r}, \quad r< A/a.
\eeq
Here the integral is supported over the resonance of width $a/A$ and height 1. At large distances
$r\gg A/a$ only the lightest modes ($m<a/A$) contribute to the exchange. 
Their coupling is unsuppressed $(\tilde{\Phi}_{m}(0)\approx 1)$ and one
can approximate

\beq\label{newtl}
V(r)\approx \int_{0}^{\infty}\frac{|\tilde{\Phi}_{m}(0)|^{2}}
{M_{*}^3}\frac{e^{-mr}}{r}dm\approx\frac{1}{M_{*}^{3}}
\frac{1}{r^2}, \quad r\gg A/a.
\eeq
This is the same behavior of Newtonian potential  as in the model of Dvali, Gabadadze and Porrati.
A weak four dimensional force is mimicked by the exchange of a continuum of massive KK gravitons
at distances smaller than the crossover distance  $r<r_c\equiv A/a$. At larger distances, 
the force becomes 
five dimensional. Crossover distance in (\ref{pc}) coincides with the 
crossover distance in \cite{dgp}. In order to
reproduce the correct Newton's constant one has to take the scale $a$ to be of the order of 
Planck scale $M_{P}$

\beq\label{rc}
r_c=\frac{A}{a}=\frac{a^2}{M_{*}^3}=\frac{M_{P}^2}{M_{*}^3}.
\eeq

At the linearized level, an exchange of massive spin
two states gives 'wrong' tensor structure of the propagator
\cite{vdvz}. It was shown \cite{dis,va}  that non-linear effects cure this problem in model \cite{dgp}. 
I conjecture that a similar mechanism occurs in the model at hand. Investigation
of that effect as well as the spectrum of induced gravity in codimensions two and three is
currently in progress \cite{pap5}.

\vspace{1cm}
{\bf Acknowledgments}
\vspace{0.1cm} \\

I would like to thank Alex Vilenkin for useful comments. I am grateful to
Gia Dvali for discussion, comments and encouraging me to make this manuscript
public.

\vspace{0.2in}
\newpage

\begin{appendix}

\section{Method for fermions}

Here we outline the procedure for finding the KK spectrum for bulk
fermion field with kinetic term that is not homogeneous in extra dimension.
The action we consider is

\beq\label{fer}
S=\int d^{5}Xf(y)\bar{\Psi}i\gamma_{A}\partial^{A}\Psi.
\eeq

Dirac algebra in 5D $\{\gamma^{A},\gamma^{B}\}=2g^{AB}$
can be satisfied with a choice of 4D gamma matrices $\gamma^{\mu}$ and 
$\gamma^{5}_{(5D)}=-i\gamma^{5}_{(4D)}\equiv -i\gamma^{5}$.

The equation of motion is independent of $f(y)$ and is solved with the ansatz

\beqa\label{azap}
\Psi(p,y)=\sum_{m}\left( g_m(y)\xi_{L}^{m}+h_m(y)\xi_{R}^{m}\right)\nn\\[2mm]
\gamma_{\mu}p^{\mu}\xi_{L}^{m}=m\xi_{R}^{m},\quad 
\gamma_{\mu}p^{\mu}\xi_{R}^{m}=m\xi_{L}^{m},
\eeqa
where spinors $\xi_{L}^{m}$ and $\xi_{R}^{m}$ together form a 
4D Dirac spinor $\psi_{m}$, i.e. $\frac{1}{2}(1-\gamma^5)\psi_{m}=\xi_{L}^{m},\quad
\frac{1}{2}(1+\gamma^5)\psi_{m}=\xi_{R}^{m}$. Wave functions are

\beqa\label{soltocop}
g_m(y)=A_m\sin(my)+B_m\cos(my)\nn\\[2mm]h_m(y)=A_m\cos(my)-B_m\sin(my).
\eeqa
Although wave functions are the same as in the case $f(y)=1$, four dimensional effective
action will be different, since one integrates over the profile $f(y)$. Plugging (\ref{soltocop})
into (\ref{fer}) and integrating over $y$ we obtain the following 4D action

\beq\label{fdaf}
S=\sum_{m,n}\left(\bar{\psi}_{m}f_{mn}i\rlap/\partial\psi_{n}+
\frac{1}{2}\bar{\psi}_{m}(m+n)f_{mn}\psi_{n} \right),
\eeq
where $f_{mn}$ is an infinite dimensional matrix given by a Fourier transform of the
profile

\beq\label{ft}
f_{mn}=\int dyf(y)\cos(m-n)y.
\eeq

Clearly, finding the spectrum of KK fermions (as well as their couplings) is now reduced to
a problem of diagonalization of infinite hermitian matrices $K\equiv f_{mn}$ and 
$M\equiv (m+n)f_{mn}/2$. Diagonalization can be done in steps, and we can write schematically

\beqa\label{diagon}
\psi^{\dagger}i\rlap/\partial K\psi+\psi^{\dagger}M\psi\longrightarrow
\psi^{\dagger}_{i}\lambda_{i}i\rlap/\partial\psi_{i}+\psi^{\dagger}UMU^{\dagger}\psi
\longrightarrow \nn\\
\psi^{\dagger}i\rlap/\partial \psi+\sum_{i,j} \psi^{\dagger}_{i}
V\frac{UMU^{\dagger}}{\sqrt{\lambda_{i}\lambda_{j}}}V^{\dagger}\psi_{j}\equiv
\sum_{m}(\bar{\psi}_{m}i\rlap/\partial \psi_{m}+m\bar{\psi}_{m}\psi_{m}),
\eeqa
where we rotated fermion sequentially as 
$\psi\longrightarrow U\psi,\psi_{i}\longrightarrow\sqrt{\lambda_{i}}\psi,
\psi\longrightarrow V\psi$. Orthogonal matrices $U,V$ are chosen to diagonalize the kinetic and
mass term. The simplest example is $f(y)=1$ in which case $f_{mn}=\delta(m-n)$ and we
recover ordinary KK tower of massive fermions. Another simple example is $f(y)=\delta(y)$ in
that case $f_{mn}$ has 1 at every entry. All eigenvalues of that matrix are zero except one that
is $2N+1$, where $N$ is (infinite) number of positive modes in $\psi$ (modes go from $-m$ to $m$).
Thus, only one mode will have nonzero kinetic term. Matrix $U$ that diagonalizes $f_{mn}$ in
this case has entries 1 on the diagonal that goes from the lower left to upper right corner,
entries -1 in first column and entries 1 in last row, while all other entries are zero.
Matrix $UMU^{\dagger}$ has zeroes on diagonal. Thus, a single propagating mode has the mass zero.
It would be interesting to find the spectrum of induced 4D fermions 
by diagonalization of the infinite matrix $f_{mn}=\int dy(1+A\exp(-(ya)^2)\cos(m-n)y$ and
compare it to the results obtained by inducing 4D kinetic term on the brane \cite{dks,dick}.

%\subsection{}

\end{appendix}

\end{document}